\documentclass[10pt]{article}
\usepackage[OE]{express}
\usepackage{amsmath,amssymb}
\usepackage{subcaption}
\usepackage{siunitx}

\def\kfrac#1#2{\leavevmode\kern.1em
\raise.5ex\hbox{$\scriptstyle #1$}\kern-.1em
/\kern-.15em\lower.25ex\hbox{$\scriptstyle #2$}}

\begin{document}
\title{Prime Number Decomposition using the Talbot Effect}

\author{Karl Pelka,\authormark{1} Jasmin Graf,\authormark{1} Thomas Mehringer,\authormark{1,2} and Joachim von Zanthier\authormark{1,2,*}}

\address{\authormark{1}Institut f\"ur Optik, Information und Photonik, Universit\"at Erlangen-N\"urnberg, 91058 Erlangen, Germany\\
\authormark{2}Erlangen Graduate School in Advanced Optical Technologies (SAOT), Universit\"at Erlangen-N\"urnberg, 91052 Erlangen, Germany}

\email{\authormark{*}joachim.vonzanthier@physik.uni-erlangen.de} 

\begin{abstract}
We report on prime number decomposition by use of the Talbot effect,  a well-known phenomenon in classical near field optics whose description is closely related to Gauss sums. The latter are a mathematical tool from number theory used to analyze the properties of prime numbers as well as to decompose composite numbers into their prime factors. We employ the well-established connection between the Talbot effect and Gauss sums to implement prime number decompositions with a novel approach, making use of the longitudinal intensity profile of the Talbot carpet. The new algorithm is experimentally verified and the limits of the approach are discussed. 
\end{abstract}


\bibliography{Talbotg1Paper}
\bibliographystyle{osajnl}


\section{Introduction}
Diffraction is one of the most fundamental aspects of optics occurring when physical objects are described by waves. The corresponding interferences may lead to surprising phenomena such as a self-imaging of periodic structures as discovered by Talbot 
\cite{Talbot} 
and theoretically derived by Rayleigh \cite{Rayleigh}.
The optical Talbot effect was later on studied theoretically in much detail, either by formulations in real space \cite{Montgomery,Liu} or phase space \cite{Banaszek,Friesch}, and experimentally observed with x-rays \cite{Cloetens}, electron beams \cite{McMorran}, atomic matter waves \cite{Chapman,Nowak} and surface plasmons \cite{Dennis,Cherukulappurath}.
Although the Talbot effect has been proven to be useful for various applications, we focus on a particular application, namely to decompose composite numbers into their prime factors. Since self-imaging is a general property of diffraction in the near field, the Talbot effect extends prime number decomposition to the field of classical optics, representing an alternative to factorizing algorithms known from quantum computation \cite{Shor97}, albeit by purely classical means. This alternative approach was initially developed by Clauser and Dowling \cite{ClauserDow} and later linked to generalized Gauss sums \cite{Berry}, which constitute a practical tool to perform prime number decompositions \cite{Woelk}. This ability of Gauss sums was already proven in many experiments e.g. by using the temporal Talbot effect \cite{Bigourd}, different interferometers \cite{Tamma2010, Tamma2011, Tamma2012, ColdAtom}, Bose-Einstein condensates \cite{BEC1} or NMR techniques \cite{NMR1, NMR2, Peng1}.
However, up to now there appears to be no experimental realization of the optical Talbot effect with the aim of performing prime number decomposition.
This circumstance becomes less surprising when noticing that complete measurements of the two-dimensional intensity profile using coherent illumination, so called Talbot carpets \cite{BerryCarpets}, were not recorded until recently \cite{Case}. 
The goal of this paper is to connect the two dots.

\section{Prime Number Decomposition using Gauss Sums \label{sec:GaussSums}}

\begin{figure}[tb]
\begin{center}
\includegraphics[width=0.92\textwidth]{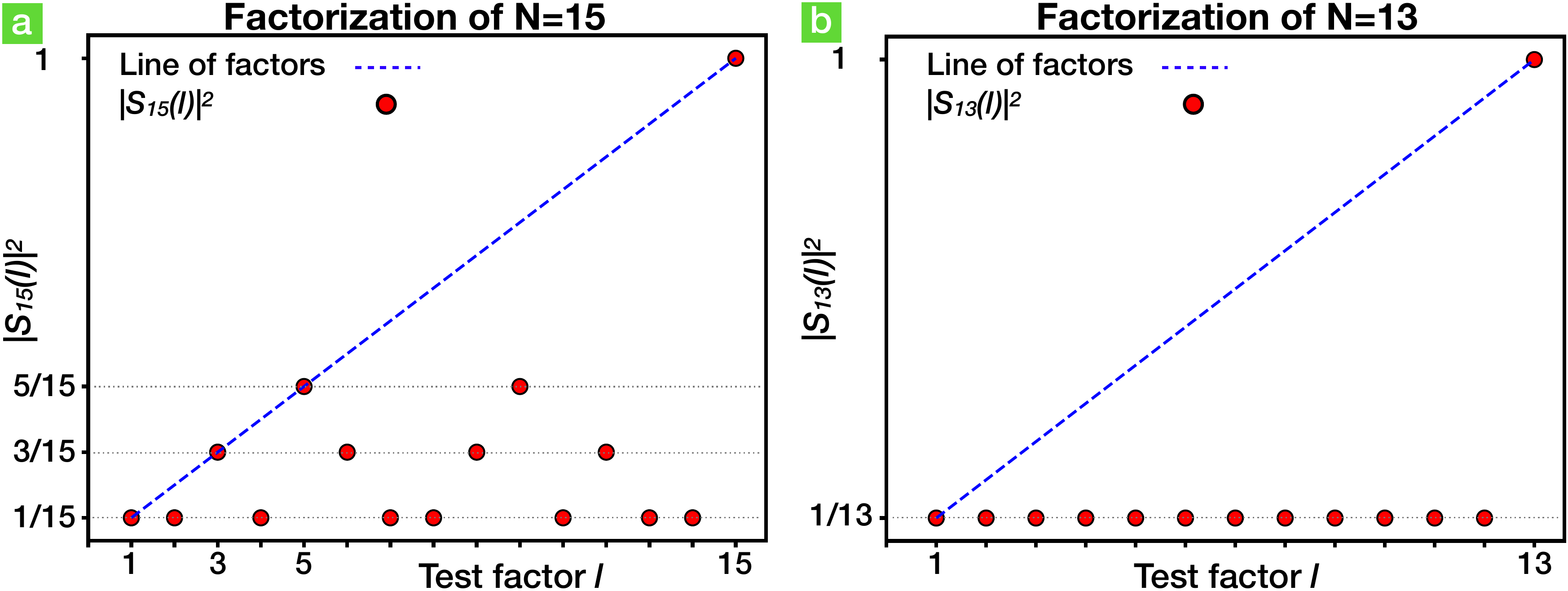}
\caption{\textbf{a} Evaluation of $|S_{15}(l)|^2$. For $l=3$ and $l=k \cdot 3$, $|S_{15}(l)|^2$ evaluates to $\kfrac{3}{15}$. For $l=5$ and $l=k \cdot 5$, $|S_{15}(l)|^2$ is equal to $\kfrac{5}{15}$. 
\textbf{b} Evaluation of $|S_{13}(l)|^2$. Here $|S_{13}(l)|^2$ evaluates to $\kfrac{1}{13}$ for every $l$ since 13 is a prime number.  Thus, all prime factors of a number $N$ intersect the line connecting $|S_N(1)|^2$ and $|S_N(N)|^2$ giving rise to the name ``line of factors''.}
\label{fig:LineOfFactors}
\end{center}
\end{figure}

Our technique to implement prime number decompositions is based on the properties of Gauss sums, which exist in many different forms \cite{Stefanak1,Stefanak2, Woelk}. 
The discrete Gauss sum, showing the characteristic property of a quadratically occurring summation index, is defined as
\begin{equation}
\label{eq:DiscrGaussSum}
S_N(l)=\sum\limits_{m=-\infty}^{\infty}w_m\exp\bigg(i2\pi m^2\frac{l}{N}\bigg),
\end{equation}
with $l,N \in \mathbb{N}$ and weight factors $w_m$, where the latter is supposed to be a slowly varying function of $m$ \cite{Woelk}. In view of the upcoming discussion, we assume the weight factor to be of the form  $w_m = a\text{ sinc}(ma) ~ \text{with } a \in \mathbb{R}^+$.
Inserting this weight factor, exploiting the periodicity of the exponential function in $m$ with period $N$ and applying the Poisson summation formula, allows us finding the convenient expression
\begin{equation}
S_N(l)=\frac{1}{N}\sum\limits_{m=0}^{N-1}\exp\bigg(i2\pi m^2\frac{l}{N}\bigg)=\frac{1}{N}G(l,N),
\label{eq:Sfunction}
\end{equation}
where $G(l,N)$ represents the so called complete Gauss sum (normalized by $N$), if the condition $a<\kfrac{2}{N}$ holds.
Employing the properties of the complete Gauss sum $G(l,N)$ we can perform prime number decomposition \cite{Woelk}.
To this aim, we assume that the number $N$ to be factorized is odd and $l$ a test factor. 
If $l$ is no factor of $N$, we obtain from Eq.~(\ref{eq:Sfunction}) $|S_N(l)|^2=\kfrac{1}{N}$, whereas if $l$ is a prime factor $p$ of $N$ or a multiple $kp$ of a prime factor, we find $|S_N(kp)|^2=\kfrac{p}{N}$. 
Hence, all prime factors of an odd number $N$ can be uniquely identified by evaluating $|S_N(l)|^2$ for all test factors $l$ and comparing with the ``line of factors $p$'', given by $f(p) = \kfrac{p}{N}$ (see Fig. \ref{fig:LineOfFactors}).


\section{The Talbot Effect \label{sec:Talbot}}

The Talbot effect is a diffraction phenomenon taking place in the near field behind a grating upon illumination with a plane wave. To derive the Talbot diffraction pattern, we assume an infinite periodic one-dimensional grating with slit width $w$ and period $d$. The corresponding transmission function reads 
\begin{equation}
T(\xi)=\sum\limits_{n=-\infty}^{\infty}\text{rect}\bigg(\frac{\xi-nd}{w}\bigg)\quad \quad\text{with} \quad \quad\text{rect}(x)=\begin{cases} 1 \hspace{1cm} \text{for} -1/2<x<1/2, \\
			    0 \hspace{1cm} \text{else}.
\end{cases}
\end{equation}
Within the Fresnel diffraction formalism we can propagate the electric field from the source plane $E(\xi,z=0)=E_0$ to the detection plane $(x,z>0)$ via 
\begin{equation}
\label{eq:ReducedFresnelIntegral}
E(x,z) = \frac{\text{exp}(ikz - i \pi /4)}{\sqrt{\lambda z}}\int\limits_{-\infty}^{\infty}T(\xi)E_0\exp\bigg\{\frac{ik}{2z}\Big[(x-\xi)^2\Big]\bigg\}\, \text{d}\xi.
\end{equation}
with $\lambda$ the wavelength and $k$ the wave number.
Since there are infinitely many slits, $T(\xi)$ is a periodic function and can be decomposed into a Fourier series 
\begin{equation}
	\label{eq:TransmissionFourierSeries}
T(\xi)=\sum\limits_{m=-\infty}^{\infty}A_m\exp\bigg(-im\frac{2\pi}{d}\xi\bigg)~~\text{with} ~~A_m=\frac{w}{d}\text{sinc}(m\frac{w}{d}).
\end{equation} 
Combining Eqs.~\eqref{eq:ReducedFresnelIntegral} and \eqref{eq:TransmissionFourierSeries} and evaluating the integral yields
\begin{equation}
E(x,z)=E_0\exp(ikz) \sum\limits_{m=-\infty}^{\infty}A_m \exp\bigg[-i2\pi\bigg(m\frac{x}{d}+m^2\frac{z}{L_T}\bigg)\bigg],
\end{equation}
where  $L_T=\kfrac{2d^2}{\lambda}$ represents the so-called Talbot length, illustrating the self-imaging effect as $E(x,0) = E(x, nL_T)$ for $n\in\mathbb{N}$.
The intensity at the grating is found by calculating the modulus square of the electric field amplitude leading to
\begin{equation}
\label{eq:TalbotEffectGauss}
I(x,z)=I_0 \bigg|\sum\limits_{m=-\infty}^{\infty}A_m \exp\bigg[-i2\pi\bigg(m\frac{x}{d}+m^2\frac{z}{L_T}\bigg)\bigg]\bigg|^2, ~~ \text{with}~~I_0=E_0^2.
\end{equation}


\section{Prime Number Decomposition using the Talbot Effect\label{sec:PrimeDecomposition}}

In 2009 a precise measurement of a Talbot carpet \cite{Case}, where also the intensity for fractional distances of the Talbot length had been determined, showed that the intensity distribution along the $x$-axis at the fractional distances $z_l \approx \kfrac{l}{N}$ resembles a Gauss sum. 
This allows to perform prime number decomposition by counting the intensity maxima parallel to the grating at the distinct distances $z_l$ \cite{Case}. 
In this paper, however, we propose a different algorithm for prime number decomposition, based on another appearance of a Gauss sum within the Talbot carpet. 
Considering the intensity distribution of Eq.~\eqref{eq:TalbotEffectGauss} at a fixed lateral position $x=qd$, $q\in\mathbb{N}$, and studying the intensity distribution along the $z$-axis, we obtain after a change of the summation index $n=-m$ and exploiting the symmetry of the Fourier coefficients $A_m=A_{-m}$ 
\begin{equation}
\label{eq:IntensityOrthogonal}
I(qd,z)=I_0 \bigg|\sum\limits_{n=-\infty}^{\infty}A_n \exp\bigg(i2\pi n^2\frac{z}{L_T}\bigg)\bigg|^2=I_0|S_{N=L_T}(l=z)|^2.
\end{equation}
Eq.~(\ref{eq:IntensityOrthogonal}) resembles already the discrete Gauss sum $S_N(l)$. 
With Fourier coefficients $A_n=(\kfrac{w}{d})\cdot \text{sinc}[m(\kfrac{w}{d})]$ it turns into a complete Gauss sum $G(l,N)$, normalized by $N$, if the condition $N<\kfrac{2d}{w}$ holds. 
This can be seen by identifying $\kfrac{w}{d}$ with $a$ in Sec. (\ref{sec:GaussSums}).  
Note that this condition limits the numbers we can factorize via the dimensions of the used grating. 
Using this expression we can rewrite Eq.~\eqref{eq:IntensityOrthogonal} as
\begin{equation}
I(qd,z)=I_0|S_{N=L_T}(l=z)|^2=\frac{I_0}{N^2}|G(l=z,N=L_T)|^2
 \quad \quad \text{for} \quad\quad N<\frac{2d}{w}.
\end{equation}
Altogether, this means that we can perform a prime number decomposition via a measurement of the Talbot effect when applying the following procedure:
\begin{itemize}
\item[1)]{Record the intensity profile of the Talbot carpet along the $z$-axis from any Talbot distance $sL_T$ to the next Talbot distance $(s+1)L_T$ at some fixed slit position $x=qd$ with $s,q \in \mathbb{N}$.}
\item[2)]{Divide the measured Talbot length equidistantly into $N$ parts by calculating $z_l=\kfrac{L_T}{N}\cdot l$ , $l = 1, \ldots, N$.  }
\item[3)]{ Draw the line of factors from $[1,I(z_1)]$ to $[N,I(z_N)]$ and locate the intensities on or above this line to find the prime factors of $N$. If $\kfrac{I(z_l)}{I(z_1)} \approx \kfrac{1}{N}$ then $l$ is no factor of $N$, if $\kfrac{I(z_l)}{I(z_1)} \approx \kfrac{l}{N}$ $l$ is a factor of $N$.}
\end{itemize}
%


\section{Experimental Realization of the Talbot Effect with Focus on Prime Number Decomposition \label{sec:ExperimentalRealization}}
\begin{figure}
\begin{center}
\includegraphics[width=0.83\textwidth]{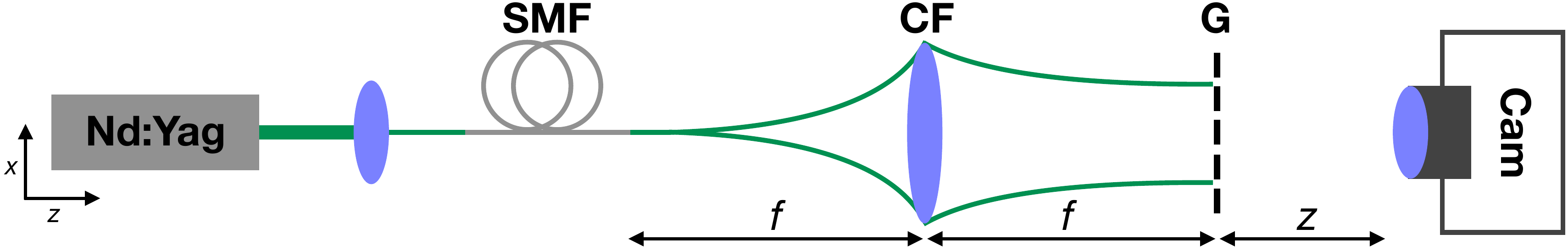}
\caption{Experimental setup: Light of a Nd:Yag laser with a wavelength $\lambda = \SI{532}{\nano\metre}$ is coupled into a single mode fiber $SMF$ in order to obtain a well-defined single spatial mode. A collimation lens $CF$ is placed at a distance of its focal length $f$ behind the fiber, such that the beam waist of the resulting Gaussian beam is located at a distance $f$ after the lens where the grating is located.
A CMOS camera equipped with a $10\times$ magnifying microscope objective is placed at a distance $z$ behind the grating. Its sensor consists of $1280 \times 1024$ quadratic pixels with a side length of $\SI{5.2}{\micro\metre}$. The camera was mounted on a translation stage which can travel $\SI{150}{\nano\metre}$ along the $z$ axis.}
\label{fig:ExperimentalSetup}
\end{center}
\end{figure}

\begin{figure}
\begin{center}
\includegraphics[width=0.94\textwidth]{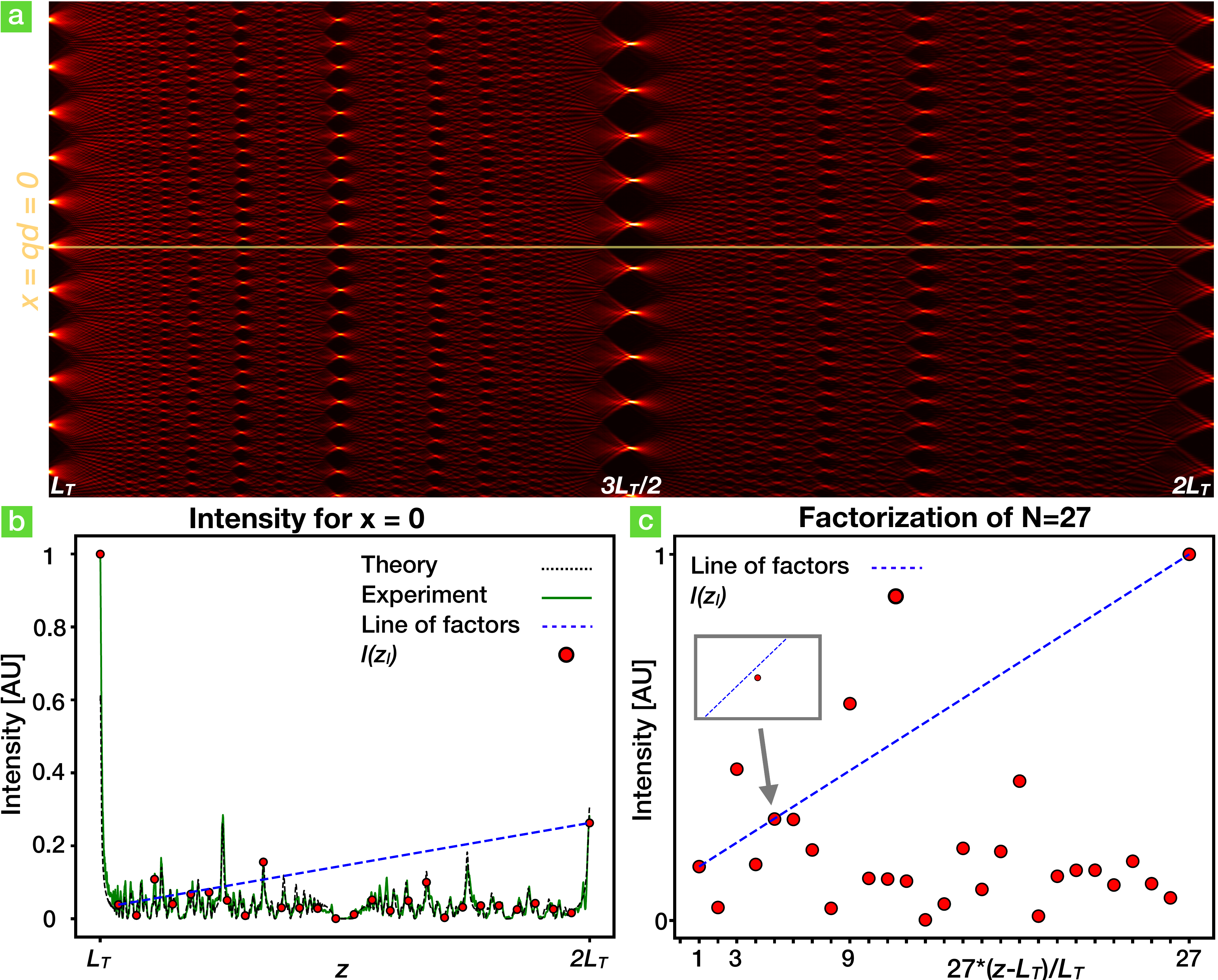}
\caption{\textbf{a} Measured Talbot carpet for a grating with period $d=\SI{200}{\micro\metre}$ and a slit width $w=\SI{10}{\micro\metre}$. The yellow line indicates the line along which the intensity is evaluated for the factorization algorithm. \textbf{b} Intensity along the yellow line as measured in the experiment (green line) and as evaluated theoretically (black dashed line) obtained by evaluation of the Fresnel diffraction integral with $250$ slits and a Gaussian beam width of $\sigma = \SI{5100}{\micro\metre}$ measured by the knife-edge method.
\textbf{c} Result of the proposed factorization algorithm using the experimental data. We see that it finds the correct prime factors $3$ and $9$ of $N=27$.}
\label{fig:Carpet}
\end{center}
\end{figure}

To prove the above proposed algorithm for prime number decomposition, we designed the experimental setup shown in Fig.~\ref{fig:ExperimentalSetup}. Placing the grating in the focus of the Gaussian beam leads to the following expression for the lateral electric field  distribution at the position of the grating
\begin{equation}
E(x,y,z=0)=E_0\exp\bigg(-\frac{x^2+y^2}{2\sigma^2}\bigg).
\end{equation}
with the Gaussian beam waist $\sigma$. 
In this case the Fresnel diffraction integral is not calculated using a plane wave as input but a Gaussian distribution, such that the intensity behind the grating reads
\begin{equation}
I(x,z)=\frac{I_0}{\lambda z}\bigg| \int\limits_{-\infty}^{\infty} T(\xi)\exp\bigg(-\frac{\xi^2}{2\sigma^2}\bigg) \exp\bigg\{\frac{i\pi}{\lambda z}\Big[(x-\xi)^2\Big]\bigg\} \text{d}\xi \bigg|^2.
\label{Fresnel2DGaussianBeam}
\end{equation}
In order to record the full Talbot carpet for a diffraction grating with a period of $d=\SI{200}{\micro\metre}$ and a slit width of $w=\SI{10}{\micro\metre}$, the measurement was automated to record with a CMOS camera images between $L_T=\SI{150.4}{\micro\metre}$ and $2L_T = \SI{300.8}{\micro\metre}$ in steps of  $\Delta z=\SI{50}{\micro\metre}$. 
From the collected data the Talbot carpet was reconstructed by stacking the measured intensity for each step, whereby each image was averaged over all rows along the $y$-axis (since the intensity does not depend on $y$).
The resulting  Talbot carpet is shown in Fig.~\ref{fig:Carpet} \textbf{a}. 
Besides the experimental data (green line) Fig. \ref{fig:Carpet} \textbf{b} displays also the theoretical prediction (black dashed line) obtained by numerical evaluation of Eq. (\ref{Fresnel2DGaussianBeam}), demonstrating an excellent agreement between theory and experiment.
As an example, Fig \ref{fig:Carpet} \textbf{c} shows the decomposition of the number $N=27$ using our new algorithm. As can be seen, the prime factors $3$ and $9$ lie above the line of factors connecting $I(z_1)$ and $I(z_{27})$ and not as before on this line. 
The reason for this is due to the reduced intensity at $z=2L_T$ compared to $z = L_T$, stemming from the Fresnel diffraction. 
Hence $I(z_{27} = 2L_T)$ lies lower than $I(z_1 = L_T)$ (see Fig. \ref{fig:Carpet} \textbf{b}) causing the slope of the line of factors to be smaller than in the ideal case. 
This effect is small, but becomes more prominent for Gaussian beams compared to plane waves. This also leads to a slight decrease of the limit up to which numbers can be factorized, theoretically given by $N < \kfrac{2d}{w}$. 
For the particular grating used, this means that we can decompose only numbers up to $N = 29$ and not as theoretically predicted up to $N=40$.
Nevertheless, the application of the discrete Gauss sum $|S_N(l)|^2$ appearing in the Talbot carpet along the $z$-direction allows us to decompose numbers within the theoretical limit of the approach into their correct prime factors. 
Our approach reduces also the amount of required experimental data to the measurement of a single Talbot carpet. The latter contains the information for all possible numbers since encoding the number to be factorized is done by merely rescaling the $z$ axis through $z_l$.


\section{Conclusion}
We theoretically discussed the capabilities of the Talbot effect to perform prime number decomposition based on its mathematical analogy to a Gauss sum.
We envisaged at first a diffraction grating with infinitely many slits illuminated by a plane wave. We then found that the properties of the ideal case are still valid for the realistic case of a diffraction grating with finite number of slits illuminated by a Gaussian beam.
The theoretical investigations result in the first realization of  the discrete Gauss sum $|S_N(l)|^2$ using the longitudinal intensity profile of the Talbot effect.
Our novel approach improves the amount of possible numbers that can be experimentally decomposed compared to existing factorization schemes based on the fractional Talbot effect. Nevertheless, we note that the Fourier coefficients of the grating transmission function yield a criterion that, depending on the geometry of the diffraction grating, limits the maximal number $N$ that can be decomposed to rather low values. However, other interference phenomena that can be described by the discrete Gauss sum $|S_N(l)|^2$ might yield less restrictive conditions and therefore enhance the possibility to implement prime number decompositions within a physical realization.

\section*{Funding}
Erlangen Graduate School in Advanced Optical Technologies (SAOT) by the German Research Foundation (DFG) in the framework of the German excellence initiative, Staedtler Foundation, Universit\"atsbund Erlangen-N\"urnberg e.V. 

\end{document}